\documentclass{vgtc}                          

\graphicspath{{figures/}{pictures/}{images/}{./}} 
\usepackage{comment}
\usepackage{times}                     
\usepackage{tabularx}
\usepackage{ragged2e}

\usepackage{array}
\newcolumntype{J}[1]{>{\justifying\let\newline\\\arraybackslash\hspace{0pt}}m{#1}}
\newcolumntype{C}[1]{>{\centering\let\newline\\\arraybackslash\hspace{0pt}}m{#1}}
\usepackage[labelfont=bf,textfont=bf]{caption}

\usepackage{tabu}                      
\usepackage{booktabs}                  
\usepackage{lipsum}                    
\usepackage{mwe}                       

\usepackage{mathptmx}                  

\onlineid{0}

\vgtccategory{Research}

\vgtcinsertpkg

\usepackage{microtype}

\title{iToT: An Interactive System for Customized Tree-of-Thought Generation}

\author{Alan Boyle
\thanks{These authors contributed equally to this work}, 
Isha Gupta$^*$, 
Sebastian Hönig$^*$, 
Lukas Mautner$^*$, 
Kenza Amara, \\
Furui Cheng,
and Mennatallah El-Assady}
\affiliation{\scriptsize Department of Computer Science\\ ETH Zürich}

\teaser{
  \centering
   \includegraphics[width=\linewidth]{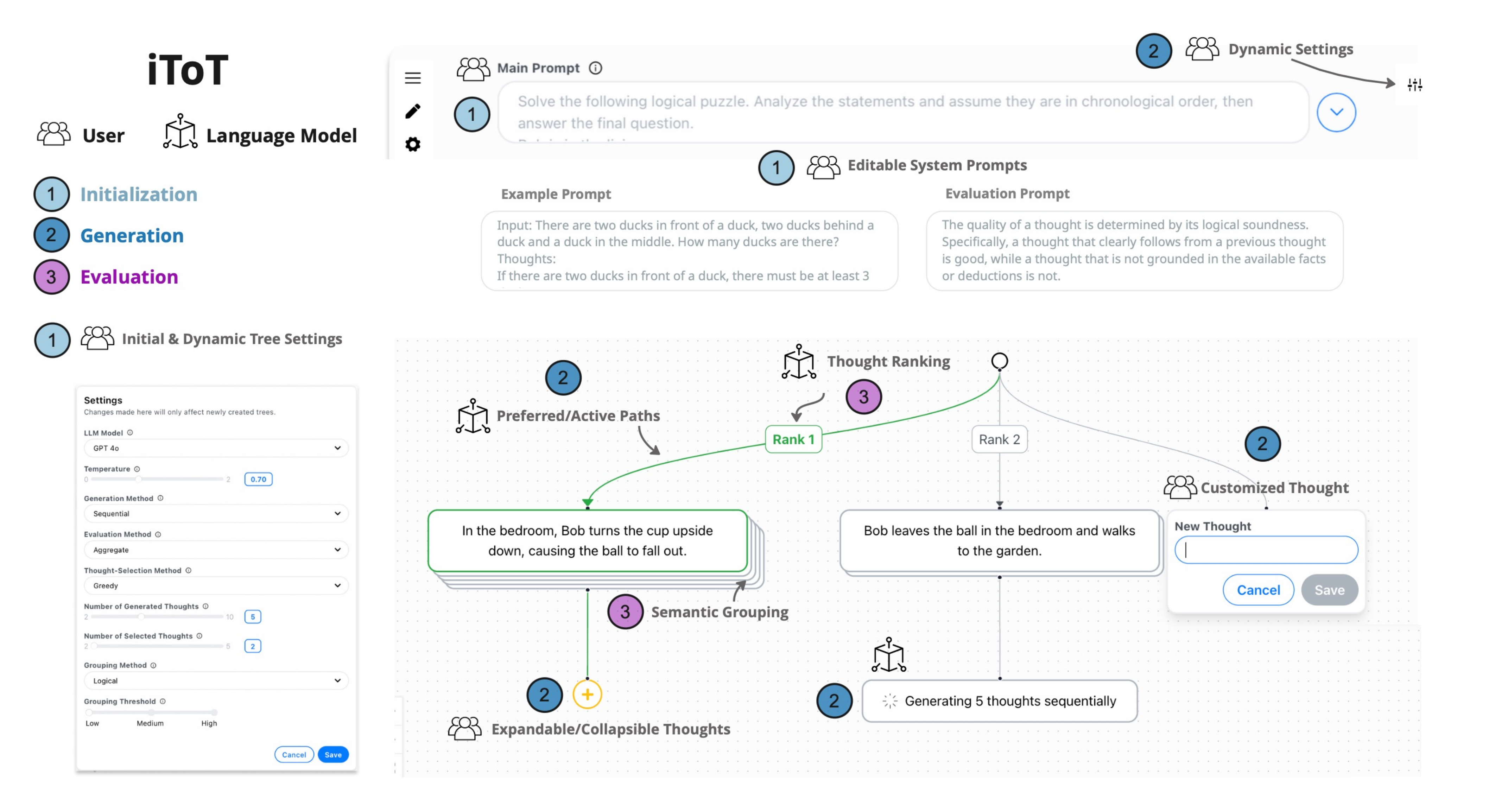}
   \caption{
        \textbf{The iToT workflow}:
        During initialization, the user provides an input prompt describing the task, examples of successful sequences of thoughts, and an evaluation prompt with self-evaluation criteria. They also specify the model parameters and visualization settings (1). During the generation process, the parametrized model produces a set of ranked candidate thoughts. The user can expand on these model-generated thoughts or add a new custom thought (2). Finally, iToT offers evaluation: thoughts are ranked by the model's self-evaluation and assessed based on their semantic similarity and self-consistency (3).
    }
   \label{fig:fig1}
 }

\abstract{
As language models have become increasingly successful at a wide array of tasks, different prompt engineering methods have been developed alongside them in order to adapt these models to new tasks.
One of them is Tree-of-Thoughts (ToT), a prompting strategy and framework for language model inference and problem-solving.
It allows the model to explore multiple solution paths and select the best course of action, producing a tree-like structure of intermediate steps (i.e., thoughts).
This method was shown to be effective for several problem types.
However, the official implementation has a high barrier to usage as it requires setup overhead and incorporates task-specific problem templates which are difficult to generalize to new problem types. It also does not allow user interaction to improve or suggest new thoughts.
We introduce iToT (interactive Tree-of-Thoughts), a generalized and interactive Tree of Thought prompting system.
iToT allows users to explore each step of the model's problem-solving process as well as to correct and extend the model's thoughts. 
iToT revolves around a visual interface that facilitates simple and \textit{generic} ToT usage and  transparentizes the problem-solving process to users.
This facilitates a better understanding of which thoughts and considerations lead to the model's final decision.
Through three case studies, we demonstrate the usefulness of iToT in different human-LLM co-writing tasks.
}


\begin{document}

\firstsection{Introduction}

\maketitle

Interest in large language models (LLMs) has grown tremendously in recent years and taken both the research community as well as the general public by storm \cite{Radford2019LanguageMA} \cite{geminiteam2024gemini} \cite{touvron2023llama}.
They have demonstrated proficiency in numerous tasks once considered extremely challenging or entirely unsuitable for machine-learning systems.
For instance, they excel in handling various logical and mathematical reasoning tasks, particularly those involving only a few logical steps to reach a conclusion \cite{plevris2023chatbots} \cite{frieder2024mathematical}.

However, despite their strengths, LLMs often struggle to answer certain types of complex questions that require planning and exploration~\cite{Hulbert}. 
Moreover, the black-box nature of LLMs complicates the possibility for humans to understand the reasoning process behind the model's decisions.

This has led to the development of prompting techniques that transparentize the LLM's inference process, like Chain-of-Thought (CoT) \cite{wei2023chainofthought}, which guides the model to emulate a step-by-step reasoning process.
CoT allows the model to iteratively build towards the correct conclusion instead of outputting it immediately, which further allows a user to analyze the model's reasoning process and identify potential shortcomings.
Tree-of-Thoughts (ToT) \cite{yao2023tree} addresses the linear limitation of the CoT approach by prompting the model to come up with multiple thoughts at each step. It thus allows the LLM to explore multiple reasoning paths simultaneously through iterative self-evaluation of the output choices.

However, ToT, as presented in the original work, is limited to specific tasks and requires the users to write code to adapt this technique to new tasks, and thus has limited usability to solve real-world problems.
The original ToT implementation runs automatically without the option for user-guided discovery and without showing the resulting tree to the user. It thus cannot adapt to user preferences nor profit from the user's knowledge and does not leverage the transparent nature of ToT as a reasoning process.
Hence, we explore how to build a general-purpose ToT system with user-guided discovery that allows the user to observe the entire reasoning process. 

Extending ToT to a general-purpose system presents several challenges. Firstly, it requires replacing the task-specific code with an intuitive task input format while still maintaining the fidelity of ToT. 
Secondly, the model may produce multiple thoughts at a given step that express the same idea, leading to duplicated options. We use the idea of semantic equivalence to reduce the elements shown to users by grouping the thoughts sharing the same idea (e.g., logically entail each other). Finally, the interface design poses a tradeoff between providing the user with sufficient information and control to interact with the system meaningfully, whilst simultaneously not overwhelming them with redundant data.

We introduce iToT (Interactive Tree-of-Thoughts). iToT allows users to input any task and interact with the model's thoughts through a web interface.
This enables transparency and contrastive reasoning: users can visualize various potential model outputs comparatively along with the quality scores assigned to each path by the model, revealing the trade-offs faced by the model and the thoughts it ultimately follows. 
Concretely, our contributions are the following:
\begin{itemize}
    \item An interactive visualization system\footnote{
The web demo is available through \url{https://itot.ivia.ch}.}, with a tree-based visualization of the ToT generation paths and natural-language-based interactions to help LLM users control and customize the ToT generation process.
    \item A semantic node grouping method to alleviate duplicated options, which makes the tree-based display more scalable with the number of branches in the generation process and also provides an indication of output consistency.
    \item Three use cases that demonstrate the usage of iToT in human-AI co-writing tasks, including problem-solving and planning.

\end{itemize}

\section{Literature Review}

\textbf{Prompting Techniques}.
Due to the ability of LLMs to solve an array of tasks going far beyond text generation, a considerable research effort into how these models can be made better at solving increasingly complex problems has been kick-started.
\textit{Prompt engineering} has proven particularly well suited for this pursuit, as it is much easier, faster, and cheaper to implement than LLM fine-tuning, and can elicit measurably improved results in many types of benchmarks.
\cite{kaddour2023challengesapplicationslargelanguage} and \cite{chen2024unleashingpotentialpromptengineering} summarize a number of such prompting strategies.
Aside from simple strategies like \textit{input-output}, \textit{impersonation} \cite{salewski2023incontextimpersonationrevealslarge} and \textit{ask-me-anything} \cite{arora2022askanythingsimplestrategy}, more and more involved techniques are being developed.
Among the most prominent of these is \textit{chain-of-thought (CoT)} \cite{wei2023chainofthought}, along with its variants like \textit{zero-shot CoT} \cite{kojima2023largelanguagemodelszeroshot} and \textit{self-consistency CoT} \cite{wang2023selfconsistencyimproveschainthought}.
All CoT-like prompting applications share the same basic idea: 
They ask the model to ``think out loud'' about the answer it is about to give.
Consequently, the model reasons about the various sub-steps required to arrive at its final answer, instead of simply generating it.
This increases the chances of a factually correct and semantically consistent response through beacons of information in the thoughts that are eventually used in the final response \cite{madaan2022textpatternseffectivechain}.
\textit{Least-to-most} prompting \cite{zhou2023leasttomostpromptingenablescomplex} has been proposed as a way to elicit complex reasoning in LLMs.
Other attempts at this have been made by \textit{self-refine} \cite{madaan2023selfrefineiterativerefinementselffeedback} and \textit{self-evaluation guided beam search} \cite{xie2023selfevaluationguidedbeamsearch}.
Especially CoT, self-consistency CoT, and self-evaluation guided beam search have served as inspiration for \textit{tree-of-thoughts (ToT)} \cite{yao2023tree}, which in turn serves as the basis for our work.
Apart from ToT, other advanced prompting strategies and frameworks have also been developed, such as \textit{graph-of-thoughts} \cite{besta2024graphofthoughts} and \textit{retrieval-augmented generation} \cite{lazaridou2022internetaugmentedlanguagemodelsfewshot, jiang2023activeretrievalaugmentedgeneration}.
Our method improves directly upon ToT by allowing users to tackle more than just the original three tasks proposed in \cite{yao2023tree} without having to write custom Python code.

\smallskip

\textbf{LLM Interfaces}.
There has also been substantial work on visualizing LLMs and their outputs.
On the technical side, \cite{bbycroftllm} offers an interface displaying the process of how an LLM works, allowing the user to interact with the model's building blocks.
Building on CoT, \cite{wu2024mindseyellmsvisualizationofthought} proposes \textit{visualization-of-thought} in an attempt to increase the spatial reasoning capabilities of LLMs and produce visualizations of their thoughts in the process.
In order to reduce the barrier to entry for users that wish to interact with LLMs in a more controlled way, \textit{Low-Code LLM} \cite{cai2024lowcodellmgraphicaluser} offers a visual programming interface that allows the construction of custom workflows.
\cite{awesomellmwebui} offers a curated list of graphical user interfaces designed for interaction with one or multiple LLMs in a purely chat-based way.
More nuanced interactions with LLMs are made possible for example by \cite{yuan2022wordcraft}, which introduces \textit{Wordcraft}, an interface that allows the user to engage in co-writing with an LLM.
Another example is \textit{Graphologue} \cite{jiang2023graphologue}, which converts an LLM's textual response into an interactive diagram, enabling the user to explore logical and associative relationships between parts of the response and display relevant sources.
Following a similar idea, \textit{RELIC} \cite{cheng2024relic} inspects the model's self-concistency and visualizes different paths within a sentence in the response using a tree-like structure, while highlighting supporting and contradicting evidence from various rounds of sampling.
These visualizations, however, all support LLM applications in which either CoT prompting or no particular prompting strategy is employed.
As of right now, no interface exists that is tailored specifically to the ToT paradigm.
To remedy this, we propose iToT.
The important difference between our method and existing, standard LLM interfaces is our display of the generated tree of thoughts, along with the produced self-evaluation rankings, instead of a standard chat interface.

\begin{figure}[h]
    \centering
    \includegraphics[width=\linewidth]{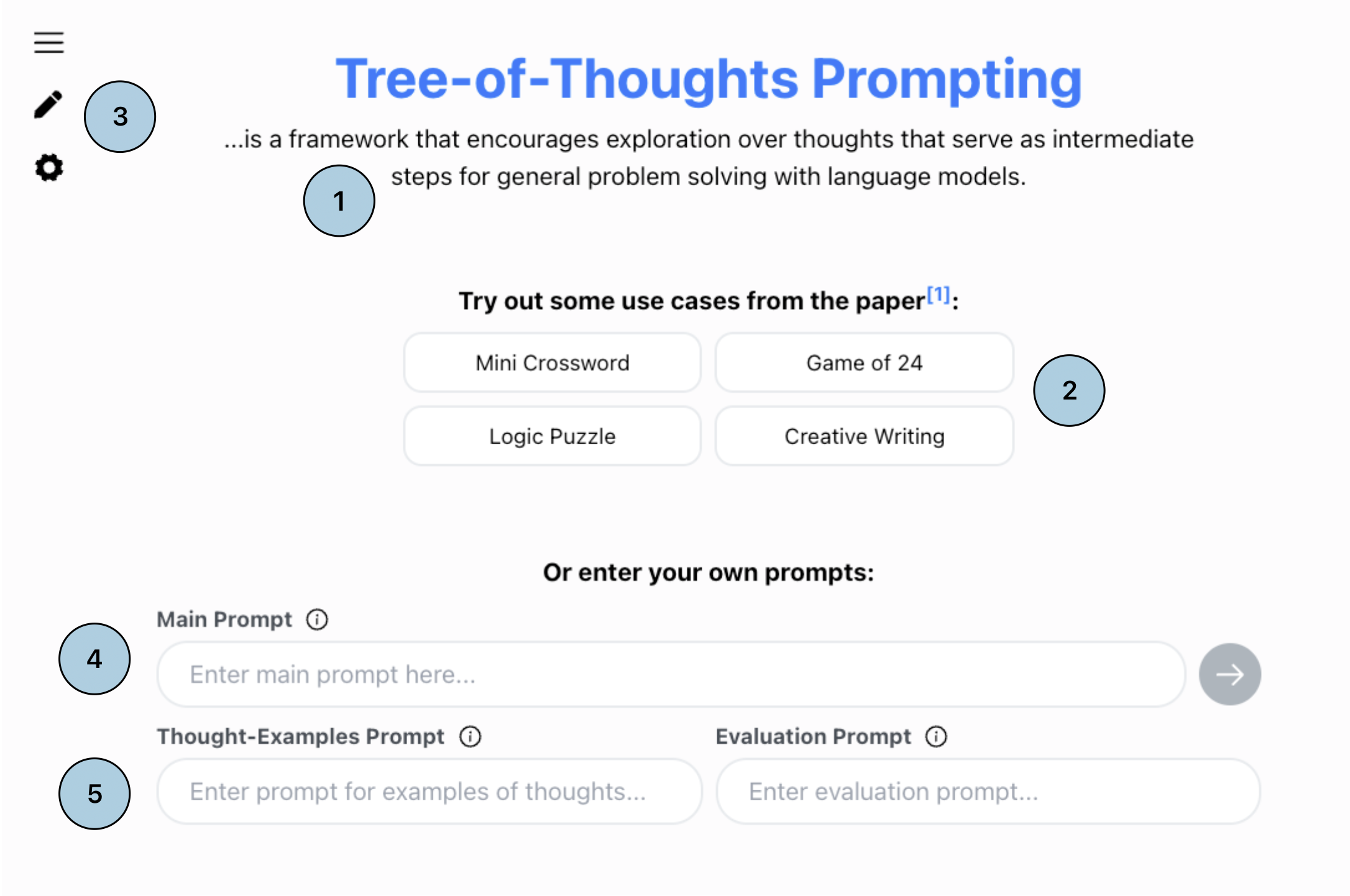}
    \caption{The landing page features onboarding guidance (1) and some example tasks for a quick start (2). The sidebar leads to the user history and tree settings (3). There is an input field for the main prompt (task description) (4) as well as the two system prompts (5). }
    \label{fig:dashboard}
\end{figure}

\section{iToT System Design}

iToT is an interactive dashboard in the form of a web application. 
The main page (shown in~\autoref{fig:dashboard}) first serves as a landing page explaining the ToT approach with some example tasks.
Once the user inputs their own prompts, the landing page is rearranged to display the resulting tree (shown in~\autoref{fig:tree}) wherein the majority of the user interaction happens. 
The sidebar offers peripheral functions such as the settings (shown in~\autoref{fig:settings}) and a history of the created trees. The layout is intended to emulate popular chatbot tools such as ChatGPT to allow users to easily orient themselves. We offer OpenAI's GPT models (GPT-3.5 Turbo \cite{gpt3}, GPT-4 \cite{gpt4} and GPT-4o \cite{gpt-4o}) to serve as the LLMs in our application. For the semantic grouping (which will be explained later) we use an SBERT \cite{reimers-2019-sentence-bert} based model as the sequence embedder for the embedding-based similarity measure and a DeBERTa \cite{DeBERTa} based NLI model for the logic-based similarity measure \footnote{The sequence embedder can be found \url{https://huggingface.co/sentence-transformers/paraphrase-MiniLM-L12-v2}. The NLI model can be found \url{https://huggingface.co/MoritzLaurer/DeBERTa-v3-base-mnli-fever-anli}.}.  

\subsection{Users and Tasks} We envision three types of users for iToT: lay users, ML engineers, and LLM researchers. Their specific needs and conceptual models were integrated into our system design. Lay users may use iToT to produce an output of better quality compared to other methods. ML engineers are interested in finding the best ToT prompting strategy for a certain task. They might want to try different combinations of prompts and ToT settings to find the best solution for the ML problem type they want to solve. LLM researchers are interested in investigating the advantages and disadvantages of the ToT prompting strategy in an open-ended manner. They may be interested in understanding all the different settings and prompt types as well as comparing the ToT output to the output resulting from a classic prompting strategy.

\subsection{ToT Interaction Overview}

To generate a ToT, the user begins by inputting the main prompt describing the task, alongside two system prompts. Before generating the tree the user can also change the settings for tree generation. This part is described in detail in Section \ref{sec:Init_Tree}.

Once the model receives the prompts it generates a set of candidate thoughts, which are then evaluated by the model and grouped according to a similarity metric by an external model. The thoughts are then displayed to the user alongside the ranking provided by the model, with grouped thoughts displayed in a stacked fashion. This process is then iterated as the user chooses a thought to be expanded and the model generates possible continuations for the path that the thought belongs to. Sections \ref{sec:Thought_Generation} and \ref{sec:Thought_Evaluation} explain these two steps of generation and evaluation in detail and also describe how they are communicated to the user.

\subsection{iToT Initialization}

\begin{figure}[ht]
    \centering \includegraphics[width=0.47\textwidth]{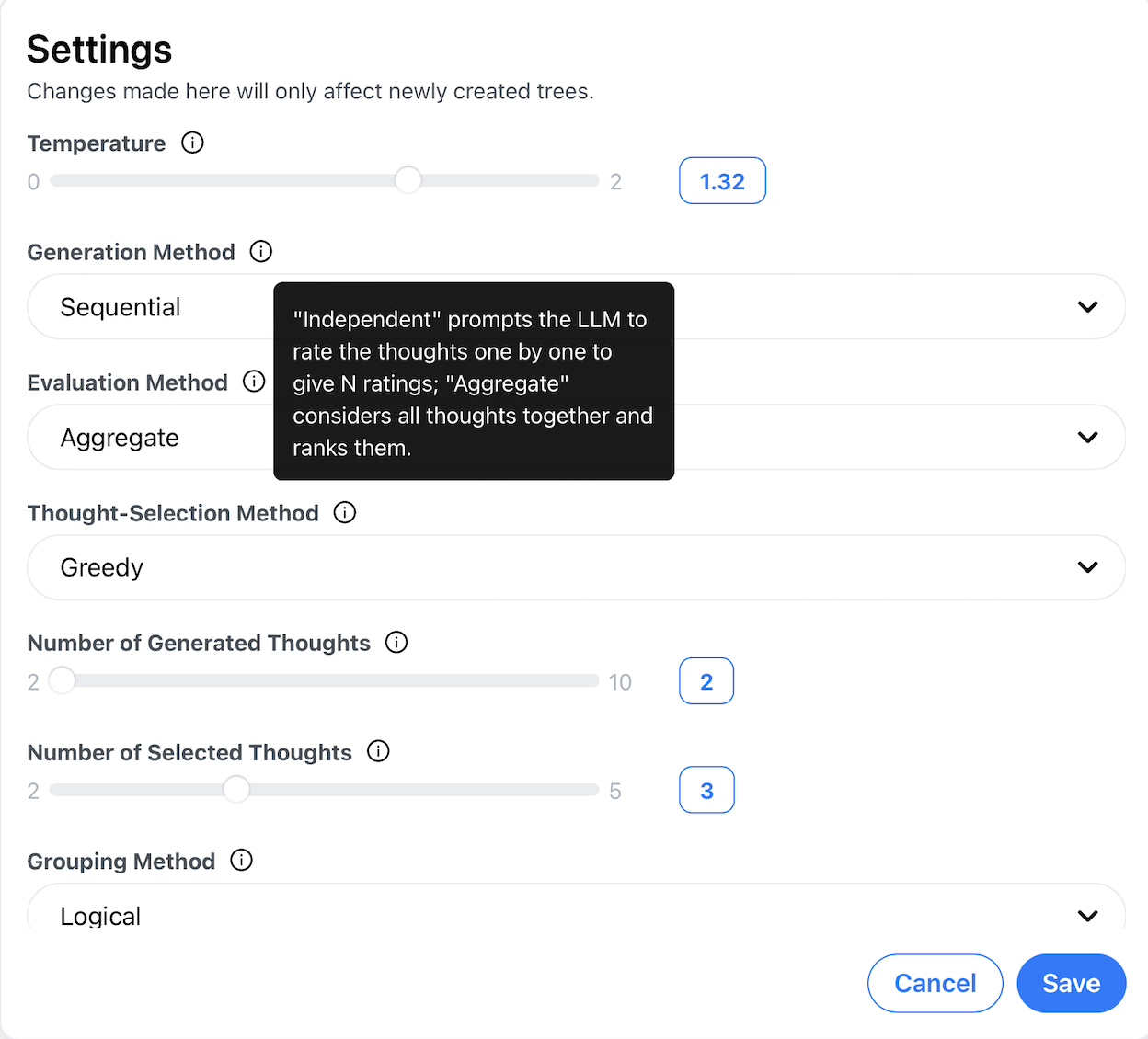}
    \caption{iToT enables users to configure the ToT process through the Setting Panel.}
    \label{fig:settings}
\end{figure}

\textbf{Initial and Dynamic Tree Settings}. 
\label{sec:Init_Tree}
The tree's initial settings (shown in~\autoref{fig:settings}) can be assigned once when initializing the tree and cannot be changed afterward. These initial settings include temperature, grouping method, generation method, evaluation method, and thought selection method. We delineate these from dynamic settings, which appear in a popup window in the top right in~\autoref{fig:tree} and can be changed at any time in the tree exploration. The dynamic settings include how many thoughts should be generated and how many of those should be displayed per layer. All settings mirror the ones that were used in \cite{yao2023tree}.

\smallskip

\textbf{Editable System Prompts}. 
The original Tree-of-Thoughts paper \cite{yao2023tree} uses task-specific prompts to 1. initialize the tree, 2. prompt the model for follow-up thoughts and 3. for the thought evaluation. We refer to these prompts as the system prompts. 
In iToT these prompts are adapted to have a fixed part and an editable part. The fixed part is the same for all inputs and sets up the model to follow the ToT approach in a generic fashion. The editable part consists of two prompts that are inserted into the fixed parts to further tailor the model's responses to the task at hand. These two prompts are:
\begin{itemize}
    \item The \textit{Example} prompt:
    This prompt shows the model examples of what a successful path in the tree could look like, akin to the examples in few-shot-prompting.
    \item The \textit{Evaluation} prompt:
    This prompt provides the criteria with which the model should evaluate its own thoughts. 
\end{itemize}
    
These prompts can be provided alongside the \textit{main} prompt which is the prompt describing the user's task, i.e., the standard input prompt users are familiar with from other applications to initialize the tree. If the user does not want to provide these editable system prompts, default alternatives are used instead, ensuring a low barrier to entry.
Through this decomposition, iToT is able to handle arbitrary input while still having the flexibility to be tailored to the task at hand.

\subsection{Thought Generation}
\label{sec:Thought_Generation}
\begin{figure*}[t]
    \centering
    \includegraphics[width=\textwidth]{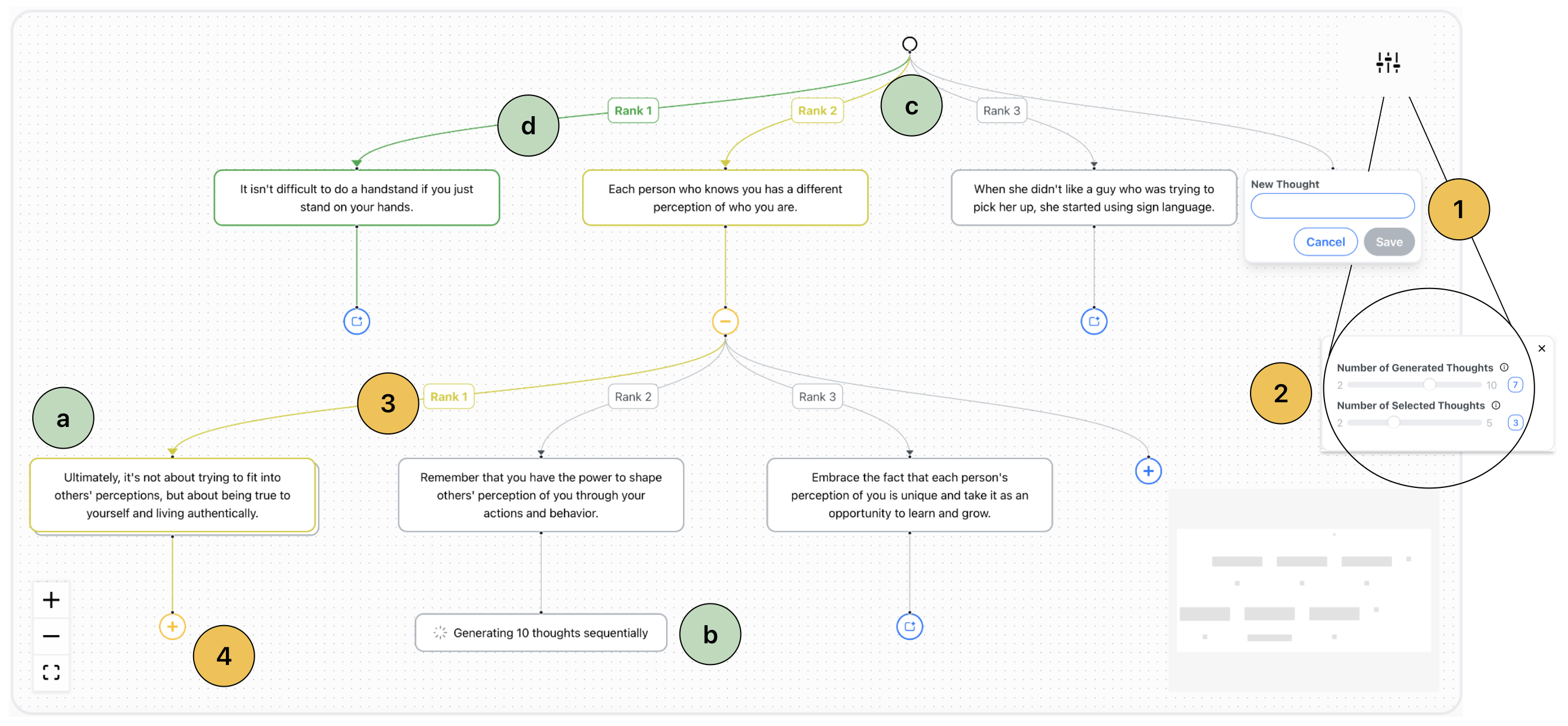}
    \caption{The user is enabled to interact with the model's thought process by adding new thoughts (1) and dynamically changing the settings for each generated layer (2). The user's "active path", i.e. the one being currently explored, is highlighted yellow (3) and they can expand and collapse subtrees for readability (4). The model supplies insights and explainability back to the user: Similar thoughts are grouped (a) and thought generation is accompanied by real-time status updates (b). The user can view the model's ranking of the generated thoughts (c) and the model's preferred path is highlighted green.}
    \label{fig:tree}
\end{figure*}

The main component of iToT is the computed Tree-of-Thought. The series of thoughts generated creates a \emph{relational} dataset.
We opt for a node-link style diagram where the vertices are arranged to depict the hierarchy of the content.
This is a natural choice as the order of thought generation must be apparent for the user to gather information from the chart.
Therefore, each node has direct lines to its $n$ children in the next layer of the tree, where $n$ is between 2 and 5 and can be chosen by the user for each layer.
The entire tree is placed within a scrollable, zoomable field so that the user can adjust the view depending on which information they are interested in.
An example of a generated tree is depicted in~\autoref{fig:tree}.
In the following, we expand on the most important design decisions of the tree component.

\textbf{Expandable/Collapsible Thoughts}. As the total number of nodes in the tree grows exponentially in the number of layers, an important tree interaction feature is the ability to hide a subset of the nodes for legibility.
When a group of children has been generated for a thought, we change the ``generate'' button to a ``collapse/expand'' toggle for all the node's children.

\textbf{Preferred/Active Paths}. We use color to identify the paths of interest in the current tree.
The path highlighted in green is the one that the model itself has evaluated as the most fruitful path to the solution -- the so-called ``preferred'' path.
The path highlighted in yellow is the one for which the user has most recently generated a new set of thoughts, which we call the ``active'' path.
This color encoding makes it easier for the user to stay focused on the important and current information in the tree.

\textbf{Addition of User's Custom Thoughts}. An important additonal feature of iToT is the ability to add a new custom thought if the user thinks that an important step is missing in the problem-solving process.
In iToT, the user can click a blue ``+'' icon to the right of any layer and add a new thought to this set of thoughts.
The children for this added thought are also immediately generated.
This allows the user to collaborate with the model to tailor the problem-solving process and arrive at the desired solution, emphasizing the mixed-initiative aspect of iToT.
It also allows the users to append thoughts that may be too original, novel, or unusual for the model to produce by itself.

\subsection{Thought Evaluation}
\label{sec:Thought_Evaluation}

\textbf{Thought Ranking}. After a set of candidate subsequent thoughts is generated for a thought, the model is asked to evaluate them either in a comparative or individual manner.
This evaluation is shown to the user alongside the thoughts as a ranking.
The user can therefore gain insights into what the model deemed to be a good or bad step.
This information is useful both for tweaking settings as well as understanding the model's thought process and limitations of the settings.
Furthermore, the system provides the option to use this evaluation when selecting the subset of generated thoughts to be displayed (known as "greedy" thought selection) - however, this selection can also be based on random sampling.

\textbf{Semantic Grouping}. We empirically noted that in some types of tasks, particularly where the answer space is fairly restricted, the model often generates several thoughts that are semantically equivalent.
For example, the two thoughts ``The astronaut's astonishment grew as he pondered the incongruity of the seared steak aroma in the weightless expanse of space'' and ``As he floated through space, he couldn't help but wonder why the smell of seared steak lingered in the vacuum''; or the two thoughts ``RILLE'' and ``Rille''.
To prevent user attention from being overwhelmed by redundant path options, we incorporate a mechanism to perform semantic grouping.
Herein, the generated thoughts are passed through a semantic equivalence filter before being displayed.
Thoughts considered equivalent by this similarity criteria are displayed in a stacked fashion, but can also be expanded manually by the user for inspection.
The grouping threshold can be changed in the settings and the user can also choose whether the grouping should happen based on similarity of sequence embeddings or logical inference.
Importantly, this feature offers explainability and supports users in understanding the semantic variations of the different branches in ToT. It indicates how sure the model is of a certain response, particularly in highly constrained tasks such as a crossword or math problem. This is further discussed in Section \ref{sec:selfcon}. \\

\subsection{User Onboarding}
An important design paradigm in interactive ML is how the users can best be informed about the current state of the system and guided in their system exploration. Several features of iToT address this.

\smallskip

\textbf{Onboarding Guidance}. We offer user guidance at the beginning of every tree creation. This includes a brief explanation of Tree-of-Thoughts prompting, a link to the associated paper, and, most prominently, four example tasks.
These tasks were taken from the official ToT repository \cite{GithubToT} as well as another ToT research project \cite{Hulbert} as exemplary problem types for which ToT is particularly suited. When selected, the three corresponding prompts and the matching selection of settings for that task are automatically inserted into the relevant input windows and settings menu respectively, so the user can easily try them out. This serves as an introduction to the dashboard and settings usage and elucidates some suitable problems.

\smallskip

\textbf{Real-time Thought Generation Status}. One notable barrier to usability within our application is the latency of the API calls to the LLM.
To alleviate the waiting time for thought generation, we include a real-time status update feature that displays the current stage of the thought generation.
This is significantly more user-friendly than the unexplained long waiting time for each new layer, which may dissuade the user from continuing usage or indeed be perceived as a bug.

\smallskip

\textbf{Tool Tips for Onboarding}. Another feature addressing user feedback is the tooltips included in various parts of the application.
By hovering over an info-icon, the user is presented with a concise explanation of its associated element.
This feature is used throughout the interface, mostly to explain ToT-specific terminology, components, or settings.

\begin{figure}[ht]
    \centering   \includegraphics[width=0.47\textwidth]{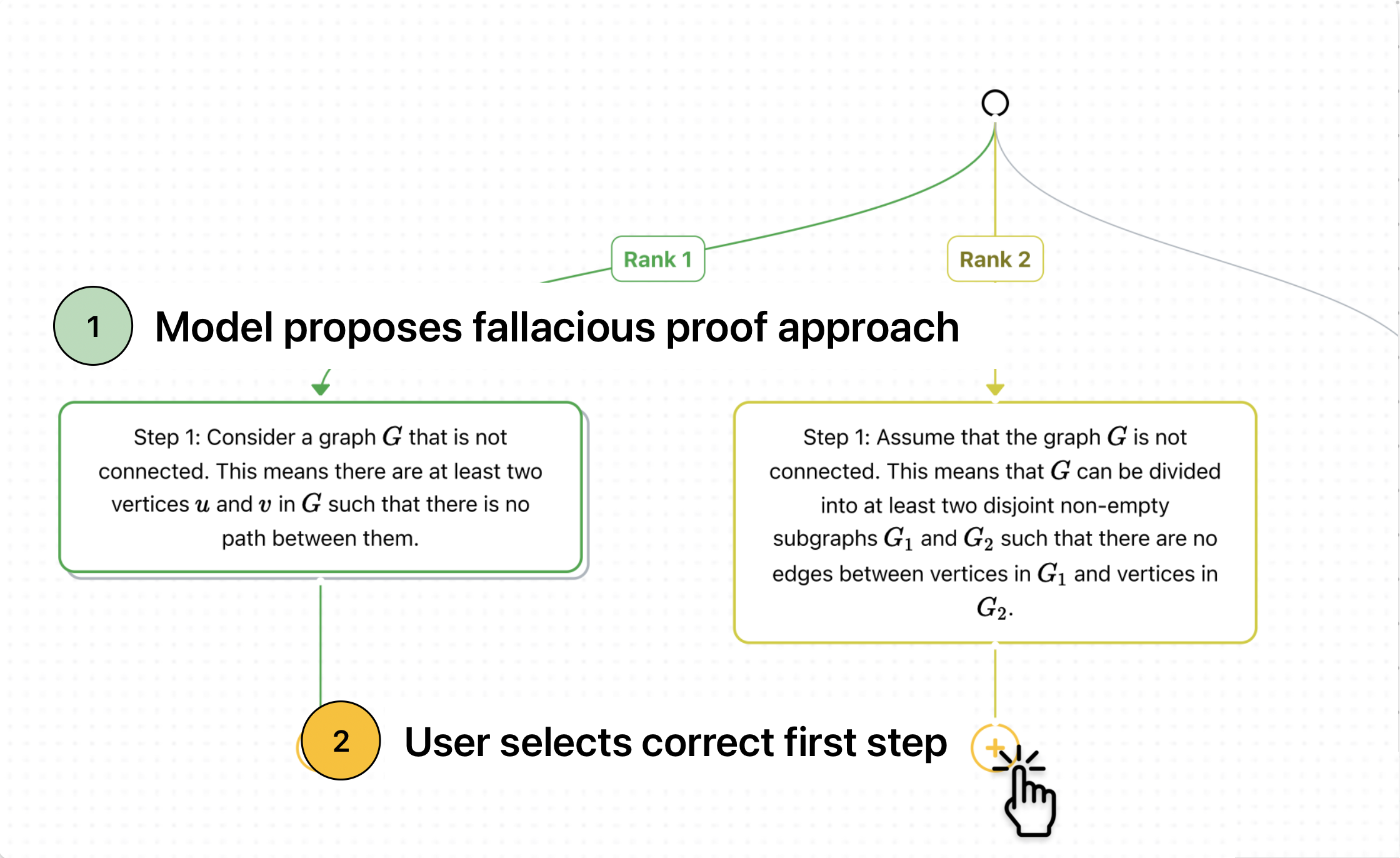}
    \caption{The first model response for the "Mathematical Proof" case study. The option evaluated as the best by the model (highlighted in green) is wrong and the option chosen by the user (highlighted in yellow) is correct. }
    \label{fig:GraphCS1}
\end{figure}

\section{Case Study}

We demonstrate how iToT achieves the aims laid out in the previous sections in three case studies. The prompts referred to here can be found in Table \ref{tab:cs_prompts} in the Supplemental Materials section.\\

\begin{figure}[ht]
    \centering   \includegraphics[width=0.47\textwidth]{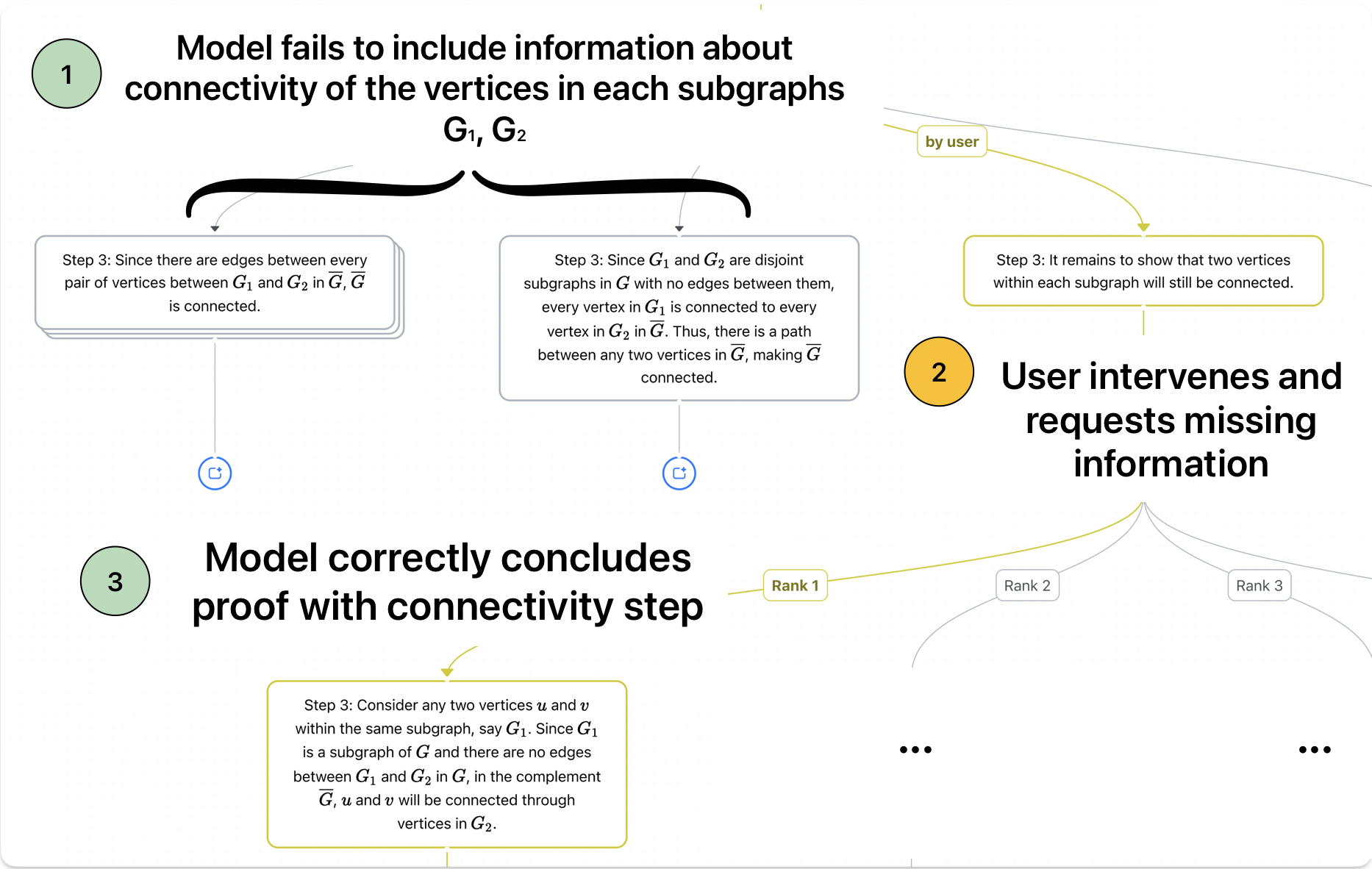}
    \caption{The user is not fully satisfied with the model's conclusion and adds a thought expressing the step the user believes to be missing. In response, the model provides a more elaborate conclusion to the proof.}
    \label{fig:GraphCS2}
\end{figure}

\begin{figure}[ht]
    \centering   \includegraphics[width=0.47\textwidth]{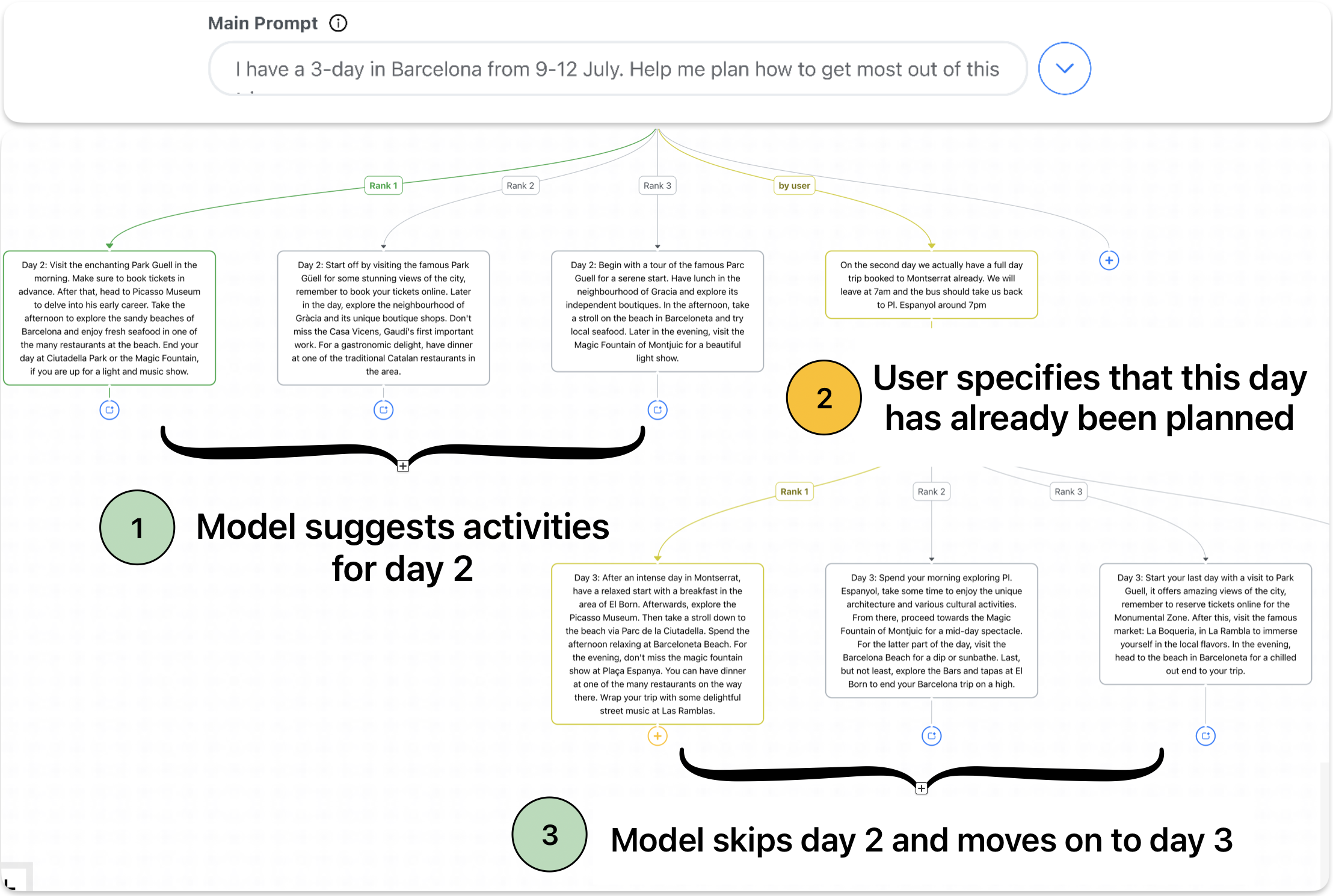}
    \caption{The user adds constraints during the generation process and the model adapts to these}
    \label{fig:holiday-study}
\end{figure}

\textbf{Case Study 1: Vacation Planning}. In this case study the user is interested in using an LLM to plan out a three-day vacation in Barcelona. This presents an open-ended task that is highly dependent on user preference.
\textit{Adapt the plan on the fly}. Alongside the main prompt asking to plan the vacation the user provides a weekend trip they had in Frankfurt as the example prompt. In the first step the model provides a diverse array of options for activities on the first day and the user agrees with the option considered best by the model. On the second day, however, the user already has a different plan which is inserted as an additional thought. The model then adapts to the user's interjection as shown in Figure \ref{fig:holiday-study} and tailors its options for the third day around the activities done on the second day.
This case study shows the merits of tailoring the model's output to the user's preferences in an open-ended setting. The ability to choose amongst a set of options instead of getting a fixed answer empowers the user to explore ideas and consider what suits them best. Moreover, the option to add thoughts can act as a fine-grained approach to alter the model's paths without needing to rerun the system from the beginning. \\

\textbf{Case Study 2: Mathematical Proof}. In this case study the user is a student working on a graph theory exercise sheet. The statement to be proven is that if a graph $G$ is not connected then its complement $\overline{G}$ is connected. Going by the definition that a graph is connected if any two vertices have a path between them the user first tried to prove the statement by looking at two vertices that do not have a path in $G$ and showing that they do in $\overline{G}$. This approach is, however, incorrect as it neglects to demonstrate that two vertices that are connected in $G$ will still be connected in $\overline{G}$. The user then prompts iToT with this problem and an example prompt that shows a step-by-step proof of the statement that the sum of degrees in a graph is even.

\smallskip

\textit{Identify the model's wrong approach}. The model provides two initial steps for the proof shown in Figure \ref{fig:GraphCS1}. The one evaluated by the system to be the best is exactly the one the user failed to prove the statement with. The user thus correctly identifies the second option as the better one. Here one can see the value of user guidance in allowing the user's knowledge and experience to enhance the ToT process, even when the user does not know the correct solution.

\smallskip

\textit{Understand the answer}. In the third step shown in Figure \ref{fig:GraphCS2} the model completes the proof, but the user is not quite satisfied as to whether this truly demonstrates that two vertices that are connected in $G$ will still be connected in $\overline{G}$. The user thus adds the thought ``It remains to show that two vertices within the same subgraph will still be connected''. The model then elaborates on this aspect giving a concrete proof of why such vertices would still be connected in $\overline{G}$, thus completing the proof to the user's satisfaction. This shows how adding thoughts can guide the model to provide better, more comprehensive answers and tailor its output to the user's understanding. In this case it was used to check whether the model is capable of arriving at a more concrete and comprehensive proof. \\

\begin{figure}[ht]
    \centering   \includegraphics[width=0.47\textwidth]{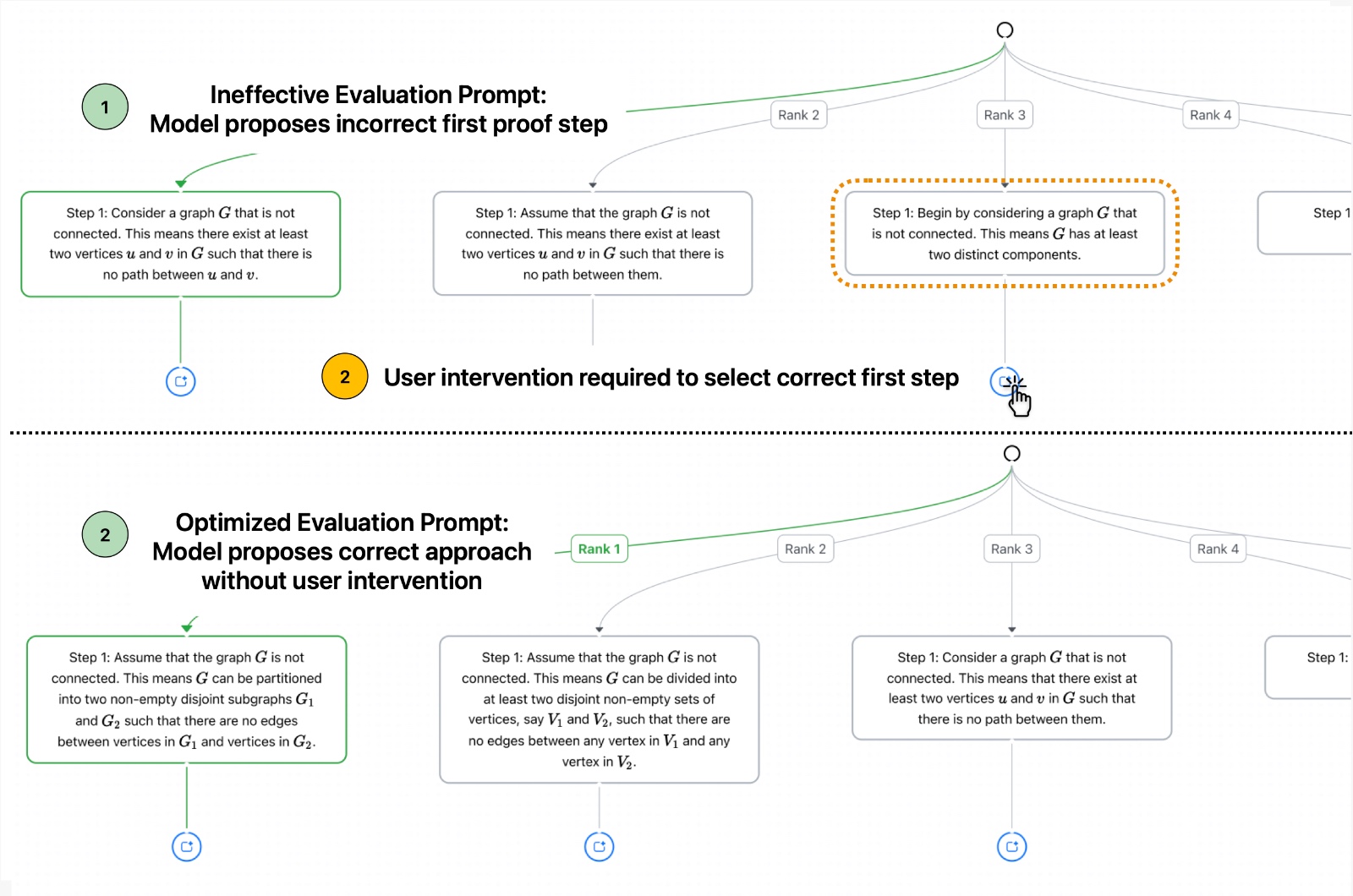}
    \caption{Evaluation prompt on top: "Steps that use more of the graph should be valued more highly". Bottom: "Steps that take a global approach as opposed to a local one should be valued more highly".}
    \label{fig:GraphCS3}
\end{figure}

\textbf{Case Study 3: Understanding the model}. In this case study, an LLM researcher is interested in gaining insights into how the given evaluation prompt shapes the model's decision making.
Building upon the example from Case Study 2, the researcher explores which evaluation prompts better enables the model to identify the correct graph-related proofs. In this particular example, the error occurred because the model focused on individual vertices rather than the whole graph. As showcased in the upper half of Figure \ref{fig:GraphCS3}, the first attempt using casual wording is unsuccessful. When phrasing this as an issue of a global vs a local approach, the model picks the correct option.
The researcher thus gains insight into the model's understanding of the graph problem by observing the correctness of the evaluation with different prompts. 
Similarly, they can compare the effect of different settings, such as temperature, on correct generation and evaluation.

\smallskip

\textit{Conclusion}. In the first two case studies we observe that our system is capable of novel tasks with little effort required from the user. This demonstrates that iToT acts as a general-purpose system. Secondly, these case studies show the need for user guidance during the generation process to profit from the user's experience and to tailor the response to the user's preferences. Lastly, the third case study shows how a researcher may use iToT to investigate how different evaluation prompts shape the model's decision making and understanding of the problem at hand.

\section{Conclusion and Discussion}

In this paper, we introduce iToT, an interactive open-domain interface designed for the custom thought generation of LLMs. To achieve more accurate, context-aware, and personalized responses, we have enhanced the interactivity at every stage of the Tree-of-Thought construction, from its parametrization to thought evaluation. By incorporating thought ranking and semantic grouping, users receive continuous feedback on the status of the ToT, allowing them to adjust their inputs accordingly. To aid user exploration, we have also included onboarding features like an initial guide and tooltips.
Our novel contribution lies in integrating user inputs at every step of the thought generation process, which we plan to further explore to improve Tree-of-Thoughts functionality. iToT offers significant benefits to various user groups: lay users can generate higher-quality text solutions for diverse problems; machine learning engineers can refine their ToT prompting strategies; and LLM experts can delve into the strengths, weaknesses, implementation of ToT, and all accompanying usage details.
As language models and chatbots become increasingly prevalent in daily tasks, iToT stands as a valuable asset in this evolving landscape.

\smallskip

\label{sec:selfcon}
\textbf{Model Self-Consistency}. Although ToT revolves around the exploration of diverse problem-solving paths, we use the semantic grouping feature to make users aware of the model's self-consistency, which can conceal issues such as hallucination and bias. This is most applicable in constrained, logical task types. For instance, if the model is answering a factual question about a named entity and none of the branches are grouped, this indicates a high variance in the model's output where it should be certain of one particular answer. The logical grouping method provides logical inference between the different answers, thus the user can use this grouping, or lack thereof, as a visual hint for clarity or contradiction within the suggested outputs.

\smallskip

\textbf{Limitations and Recommended Usage}. We notice that the quality of the output depends heavily on the presence and suitability of the two system prompts.
We strongly recommend changing them depending on the task at hand.
In case the default system prompts are used, we observe good results for more creative tasks but a worse performance for tasks with strict and clear constraints, such as the crossword puzzle. Within the iToT framework, we offer GPT-3.5 Turbo, GPT-4, and GPT-4o. Although GPT-3.5 Turbo API calls are considerably less expensive, we note a higher output quality - particularly in adherence to the desired output structure - when working with the newer models and thus suggest using the GPT-4 series for the best results.

\smallskip

\textbf{Future Work}. To expand the iToT framework, we propose:
\begin{itemize}
    \item \textbf{Automatic Tree Extension.} Currently, each layer of the tree has to be generated through active user decisions and actions. The natural next step is to have an auto-solve mode, wherein the full path to the best final answer is generated automatically according to the model's self-evaluation. 
    \item \textbf{Custom Output Parsing:} Generated thoughts in iToT could follow a specific format, for example, using a regular expression so that the user can extract only the necessary solution from each generated thought. This would help in constrained problems such as the crossword puzzle.
    \item \textbf{Token-Level Tree-of-Thoughts:} 
    Limiting for the number of tokens generated per thought for more fine-grained control.
    \item \textbf{Explainability:} iToT could offer more transparency to users to grasp which parts of the main prompt resulted in a particular word in one of the output thoughts.
    
\end{itemize}

\bibliographystyle{abbrv-doi}

\bibliography{template}

\newpage
\newpage
\onecolumn
\section*{Supplemental Materials}
\label{sec:supplemental_materials}
\subsection*{iToT technical stack}
iToT consists of a backend and a frontend, communicating with each other via HTTP requests.
Specifically, our backend serves a RESTful API, allowing the frontend to perform the actions requested by the user.
Our backend is written entirely in Python, using the FastAPI library\footnote{\href{https://fastapi.tiangolo.com/}{https://fastapi.tiangolo.com/}} to serve the previously mentioned API.
We communicate with these models using a Microsoft Azure subscription and the relevant Python packages.
We serve the backend using Uvicorn\footnote{\href{https://www.uvicorn.org/}{https://www.uvicorn.org/}}.
Our frontend is written in React\footnote{\href{https://react.dev/}{https://react.dev/}} and served via Vite\footnote{\href{https://vitejs.dev/}{https://vitejs.dev/}}.

\subsection*{Case Study Prompts}

\renewcommand{\arraystretch}{1.7}

\begin{table*}[ht]
\centering
\begin{tabular}{ |C{0.9in}|J{1.3in}|J{2in}|J{1.5in}|  }
 \hline
  \Large \textbf{Case} &\multicolumn{3}{c|}{\Large \textbf{Prompts}} \\
  \cline{2-4}
 \Large \textbf{Studies} &  \large \textbf{Main Prompt} & \large\textbf{Example Prompt} & \large \textbf{Evaluation Prompt} \\
 \hline

 \hline
 \large \textbf{Vacation Planning}   & I have a 3-day in Barcelona from 9-12 July. Help me plan how to get the most out of this trip.   &Input: Help me plan a weekend in Frankfurt. \break
Day 1: Visit the Dom/Römer area and enjoy a cozy walk in Oldtown. Make sure you walk across the main and if the weather is good even try stand-up paddling. \break
Day 2: Try out the famous Apfelwein (Äppler) in the old Sachenhaus district. If you're into shopping then visit the Zeil.
&   The quality of a thought is determined by its coherence with the thoughts in the chain before it and its contribution to solving the problem at hand.\\
 \hline
  \large \textbf{Mathematical Proof (User)}   & Prove that if a graph is not connected then its complement is connected.    &Input: Show that the sum of all degrees of a graph is even. \break
Step 1: Take the sum over all degrees. \break
Step 2: Notice that this some counts every edge in the graph twice.\break
Step 3: Thus, this sum is two times the number of edges in the graph. \break
Step 4: Hence the sum of all degrees is even.&   The quality of a thought is determined by its coherence with the thoughts in the chain before it and its contribution to solving the problem at hand.\\
\hline
 \large \textbf{Mathematical Proof (Researcher Unsuccessful) }   & Prove that if a graph is not connected then its complement is connected.    &Input: Show that the sum of all degrees of a graph is even. \break
Step 1: Take the sum over all degrees. \break
Step 2: Notice that this some counts every edge in the graph twice.\break
Step 3: Thus, this sum is two times the number of edges in the graph. \break
Step 4: Hence the sum of all degrees is even.&   Steps that use more of the graph should be valued more highly.
\\
\hline
 \large \textbf{Mathematical Proof (Researcher Successful)}   & Prove that if a graph is not connected then its complement is connected.    &Input: Show that the sum of all degrees of a graph is even. \break
Step 1: Take the sum over all degrees. \break
Step 2: Notice that this some counts every edge in the graph twice.\break
Step 3: Thus, this sum is two times the number of edges in the graph. \break
Step 4: Hence the sum of all degrees is even.& Steps that take a global approach as opposed to a local one should be valued more highly.
\\
\hline
\end{tabular}
\caption{The table shows the prompts used in the case studies.}
\label{tab:cs_prompts}
    
\end{table*}

\end{document}